\documentclass[journal]{IEEEtran}
\bstctlcite{IEEEexample:BSTcontrol}
\usepackage{ragged2e}
\usepackage{array}
\usepackage{amssymb}
\usepackage{cite}
\usepackage{graphicx}
\usepackage{hyperref} %<--- Load after everything else
\usepackage[font=small,labelfont=bf,justification=justified]{caption}

\begin{document}
\title{\huge{\textbf{Experimental demonstration of bandwidth enhancement in photonic time delay reservoir computing}
}}

\author{Irene Est\'ebanez$^{*}$, Apostolos Argyris, and Ingo Fischer  % <-this % stops a space

\thanks{I. Est\'ebanez, A. Argyris, and I. Fischer are with Instituto de Física
Interdisciplinar y Sistemas Complejos IFISC (CSIC-UIB), Campus UIB,
Palma de Mallorca, 07122, Spain (e-mail: irene@ifisc.uib-csic.es;
apostolos@ifisc.uib-csic.es; ingo@ifisc.uib-csic.es).}}

\maketitle
\begin{abstract}
\normalsize{Time delay reservoir computing (TDRC) using semiconductor lasers (SLs) has proven to be a promising photonic analog approach for information processing. 
One appealing property is that SLs subject to delayed optical feedback and external optical injection, allow tuning the response bandwidth by changing the level of optical injection. 
Here we use strong optical injection, thereby expanding the SL's modulation response up to tens of GHz.
Performing a nonlinear time series prediction task, we demonstrate experimentally that for appropriate operating conditions, our TDRC system can operate with sampling times as small as 11.72 ps, without sacrificing computational performance.}
\end{abstract}

\IEEEpeerreviewmaketitle

\section{Introduction} \label{sec:intro}
Photonic systems that perform analog information processing have been demonstrated in recent years as an interesting alternative to conventional digital computing. 
In particular, the use of photonic devices in brain-inspired computing and machine-learning schemes has attracted significant attention, helping to reduce learning costs and power consumption \cite{shastri2021photonics}. 
Among the different techniques, reservoir computing (RC) \cite{jaeger2004harnessing,tanaka2019recent,Nakajima2021} has proven to be a powerful method, drastically simplifying the implementation and training of recurrent neural networks. 
The time-delay reservoir computing (TDRC) approach \cite{appeltant2011information,larger2012photonic,paquot2012optoelectronic,brunner2013parallel} represents a very successful minimal design approach of RC.
TDRC uses time-multiplexing implemented via temporal masking to recurrently connect virtual nodes in a delayed feedback loop.
This allows the storage of past information and the generation of different responses depending on the previous inputs. 
While the recurrence is established by the feedback loop (time delay, $\tau$), the coupling among virtual nodes can be introduced via a mismatch of delay $\tau$ and masking period $T_{m}$, and via the inertia of the transient response of the real node. 
For achieving coupling through inertia, the separation between the virtual nodes $\theta$ must be smaller than the response time of the nonlinear node $T$, but not too small, to obtain an acceptable level of signal-to-noise ratio (SNR) responses. 
The masked information is introduced into the reservoir by an external drive laser. 
For this drive-response configuration, we show that the reservoir's response bandwidth can be increased by the high power of the injected optical carrier. 

In this letter, we experimentally demonstrate that a significantly higher processing speed of a photonic TDRC can be achieved by exploiting the bandwidth enhancement of the response SL. 
Specifically, under a strong optical injection of the input information, we reduce the virtual node separation to only 11.72 ps - the smallest reported so far - while preserving the TDRC's computational performance obtained for larger separations (e.g. 93.75 ps). 
This work complements and confirms our previously published numerical results \cite{estebanez2020accelerating}.

\section{Experimental realization}  
\label{sec:setup}

\subsection{Photonic reservoir}
\begin{figure}[h]
\centering\includegraphics[width=0.85\columnwidth]{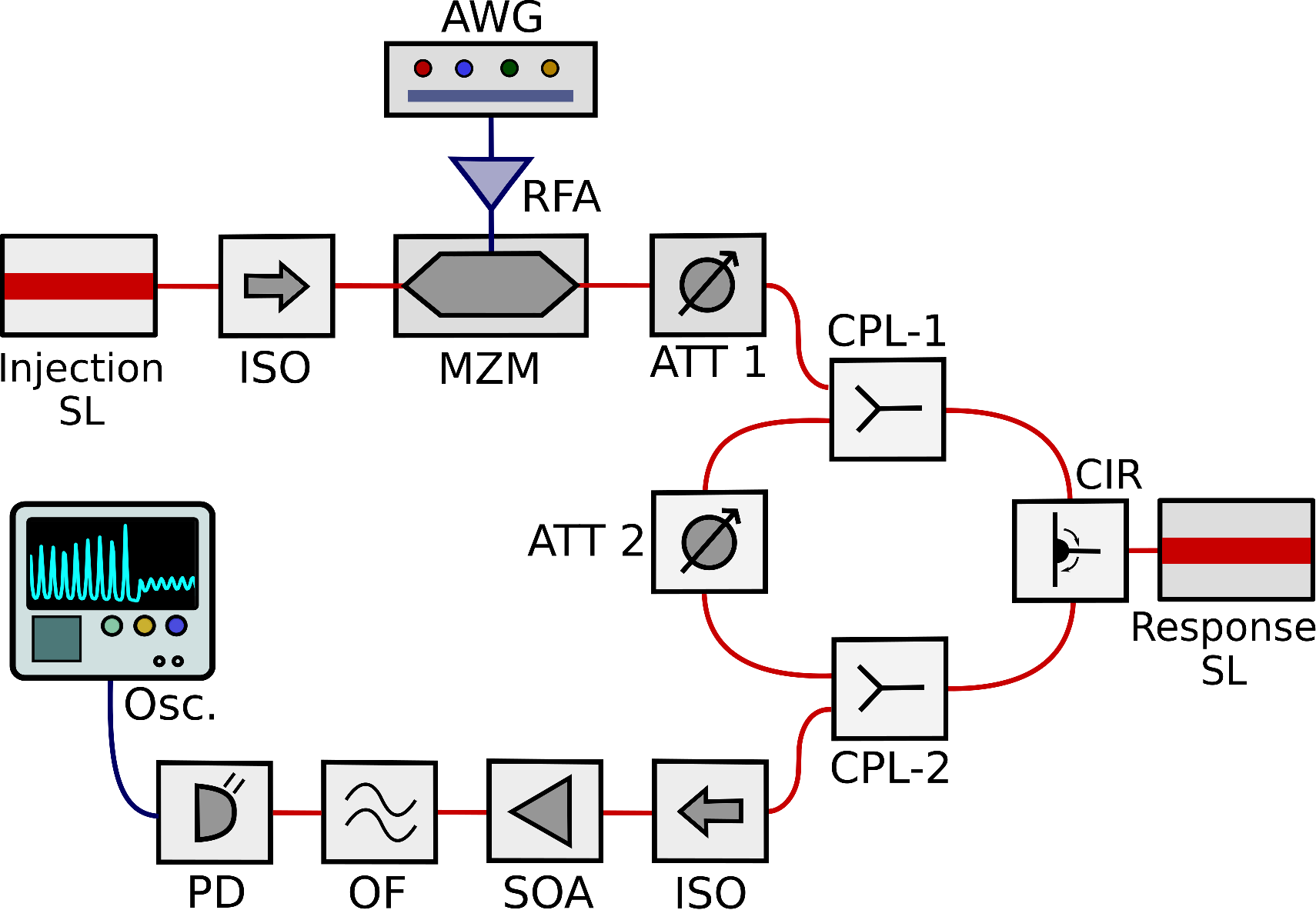}
\caption{Experimental photonic TDRC. ISO: optical isolator.}
\label{fig:setup}
\end{figure}

\begin{figure}[h]
\centering\includegraphics[width=0.65\columnwidth]{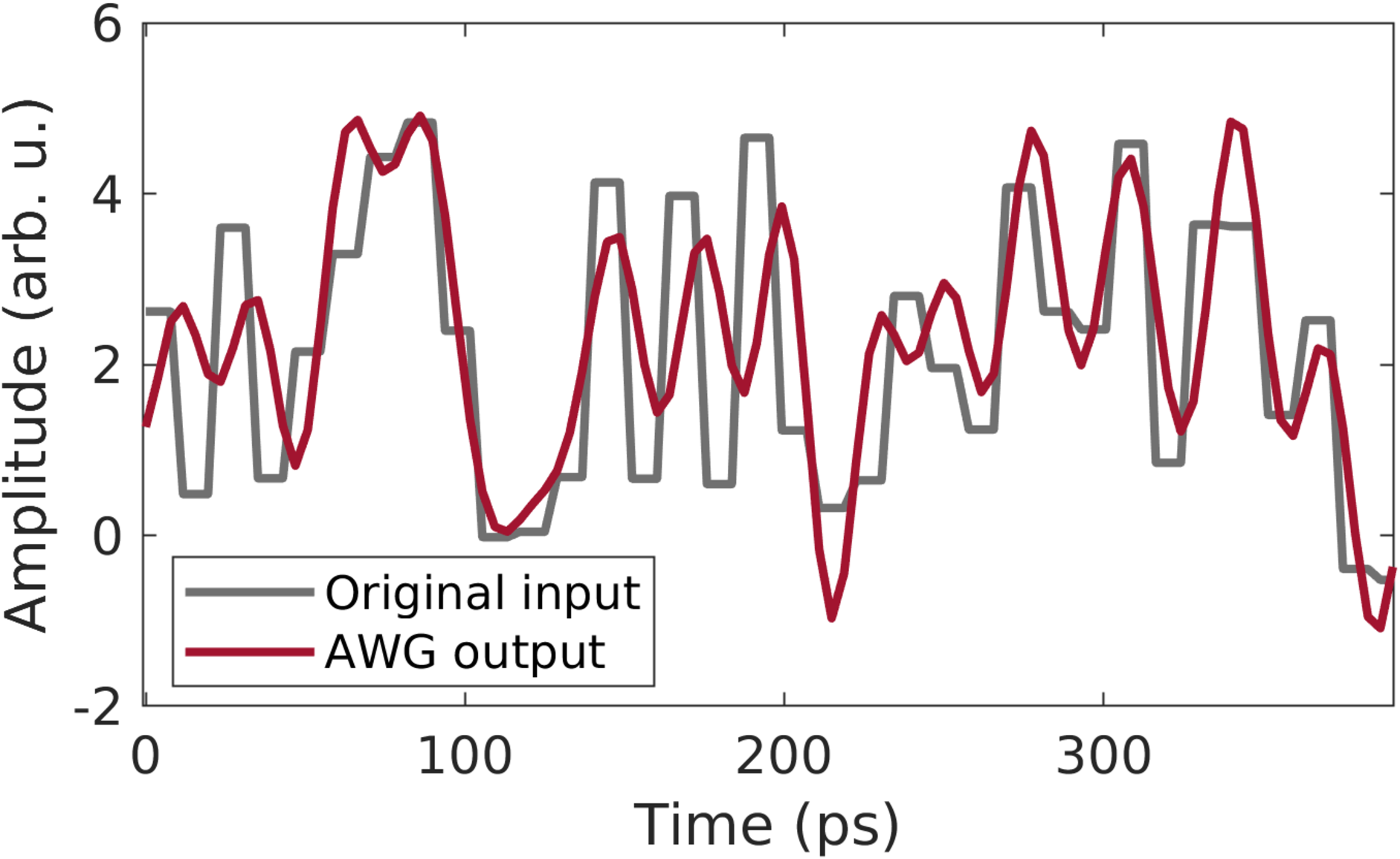}
\caption{Pre-uploaded masking sequence to the AWG (gray) and RF bandwidth-limited generated masking sequence by the AWG (red).}
\label{fig:awgres}
\end{figure}

The experimental single-mode fiber-based (SMF) photonic reservoir is shown in Fig. \ref{fig:setup}. 
The input signal is generated by multiplying each value of the to-be-processed information sequence with a mask to expand its dimensionality. 
This mask is a periodically repeated sequence of length $T_m$, drawn randomly from a uniform distribution $[0, 1]$. 
The masked input signal is uploaded into an arbitrary waveform generator (AWG - Keysight M8196A, 92 GSa/s, 32 GHz) and transformed into an electrical modulation signal with a sampling rate of 85.33 GSa/s.
Each value is assigned to only one sample, resulting in an analog bandwidth-limited AWG output signal (Fig. \ref{fig:awgres}).
The analog bandwidth limitation causes some additional correlations between virtual nodes, which can be beneficial to computing tasks.  

The AWG’s output is amplified with a 55 GHz SHF-S807C broadband RF amplifier (RFA) and modulates the optical carrier of an injection DFB laser (SL), via a 40 GHz Mach-Zehnder intensity modulator (MZM - iXblue MX-LN-40 with $V_\pi$ = 5.3V) that operates in the linear regime. 
The resulting optical signal is injected into a response DFB laser via a 50/50 optical coupler (CPL-1) and an optical circulator (CIR), while its strength is controlled by an optical attenuator (ATT 1).
The photonic reservoir is implemented using a fiber loop with a roundtrip delay of $\tau=24.5$ ns. 
The setup is realized using polarization-maintaining (PM) SMF and components, ensuring robust operation over time. 
The response laser is emitting at 1545.5 nm and is biased below threshold at 10.6 mA ($I_{th}=10.8$ mA).
The drive SL emits at a similar wavelength that can be tuned via temperature control. 
The frequency detuning between the drive and the response laser $\Delta f = f_{d}- f_{r}$ can be changed with a resolution of 0.01 K ($\sim$ 125 MHz) and is a crucial control parameter since it determines the dynamical response to the input sequence. 
A 10 dB optical attenuator (ATT 2) sets the optical feedback strength, and a 50/50 optical coupler (CPL-2) closes the feedback loop. 
The optical output of the response laser is amplified by a semiconductor optical amplifier (SOA), filtered by a tunable optical filter (OF), and detected by a 40 GHz photoreceiver (PD). The converted electrical signal is obtained via a real-time oscilloscope (Osc - Keysight UXR0404A, 256 GSa/s, 40 GHz). 
Finally, the TDRC's virtual node responses obtained from recorded time series - obtained after 256 averages for SNR improvement - are used to train a linear (LR) classifier and to evaluate the computing task performance.

\subsection{TDRC and benchmark task}

Given the feedback delay $\tau = 24.5$ ns of our photonic reservoir (Fig. \ref{fig:setup}), the 85.33 GSa/s sampling rate sets the space for 2090 virtual nodes in the TDRC. 
We use 2080 of the available virtual nodes. 
Thus, the mask length is asynchronous to the delay ($T_{m}\neq\tau$), increasing the connectivity in the reservoir \cite{stelzer2020performance}. 
We investigate two different cases of virtual node separation. 
For the smaller separation ($\theta_s = 11.72$ ps), every encoded input information is masked with a random value at the sampling rate of the AWG.
For the larger separation ($\theta_l = 93.75$ ps), every encoded input information is masked with a random value that is repeated eight times, setting an effective sampling rate for processing to 10.67 GSa/s. 
This results in a TDRC with 2080/8 = 260 virtual nodes.  
For a fair comparison of the TDRC performance, we implement the same number of virtual nodes ($N=260$) also when using $\theta_s$, by repeating the encoded masked input sequence eight times to fill the delay $\tau$.

We evaluate the TDRC's performance via a benchmark test that has commonly been used in the reservoir computing community, the Santa Fe time series prediction \cite{weigend1993results}. 
The aim is to predict the future value of a chaotic time series, $y(t+1)$, by considering its previous values up to the time $t$. 
At the output layer of the TDRC, we consider the responses from the first 3000 points of the Santa Fe time series to train the system via an offline, ridge regression algorithm with ridge parameter $\lambda = 0.01$. 
We discard the next 500 points to eliminate prediction bias, and we apply the calculated weights from the training process to a test set of the next 1000 data points. 
We use the normalized mean square error (NMSE) to quantify the prediction performance:
\begin{equation}
NMSE = \frac{1}{L} \sum^{L}_{n=1} [y(n)-\bar{y}(n)]^{2}
\end{equation}
where $L$ is the number of data points used in the test set. $y$ is the predicted value, $\bar{y}$ is the expected value, and are both normalized to zero mean and unit variance. 

\section{Results}
\label{sec:Res}
The dynamical response of the TDRC defines the nonlinear transformation of the input signal, which determines the computing performance.
One attribute of the dynamical response is the TDRC's operating bandwidth, which has to be sufficient to generate large enough response signals despite small $\theta$
%($\sim 0.2$ T) 
\cite{ortin2020delay}. 
To obtain the best computing performance, additional conditions must also be fulfilled; attributes, such as fading memory and consistency of the nonlinear input-output transformation, are critical for the final performance \cite{bueno2017conditions}. 
In the following, we evaluate the effect of the bandwidth-enhanced operation on the Santa Fe prediction task for several frequency detuning conditions ($\Delta f$). 
We identify those conditions that result in bandwidth-enhanced operation, and we show how we can benefit from faster transient states and smaller $\theta$. 

\subsection{Bandwidth enhancement} % or response bandwidth

It is known that in injection-locked SLs with strong optical injection, the response bandwidth can be several times the free-running relaxation oscillation bandwidth \cite{simpson1995bandwidth,liu1997modulation,murakami2003cavity,wang2008enhancing,kanno2016complexity}. 
To measure the bandwidth enhancement of our system, we upload random values chosen from a uniform distribution to the AWG, and choose 85.33 GSa/s as the output sampling rate.
The optical output is averaged 2048 times before calculating the corresponding spectra. 
We define the reservoir’s response bandwidth as the frequency where a 10 dB power spectral density reduction occurs, with respect to the lowest frequency components of the time series we capture in the MHz regime.

\begin{figure}[t]
\centering\includegraphics[width=0.95\columnwidth]{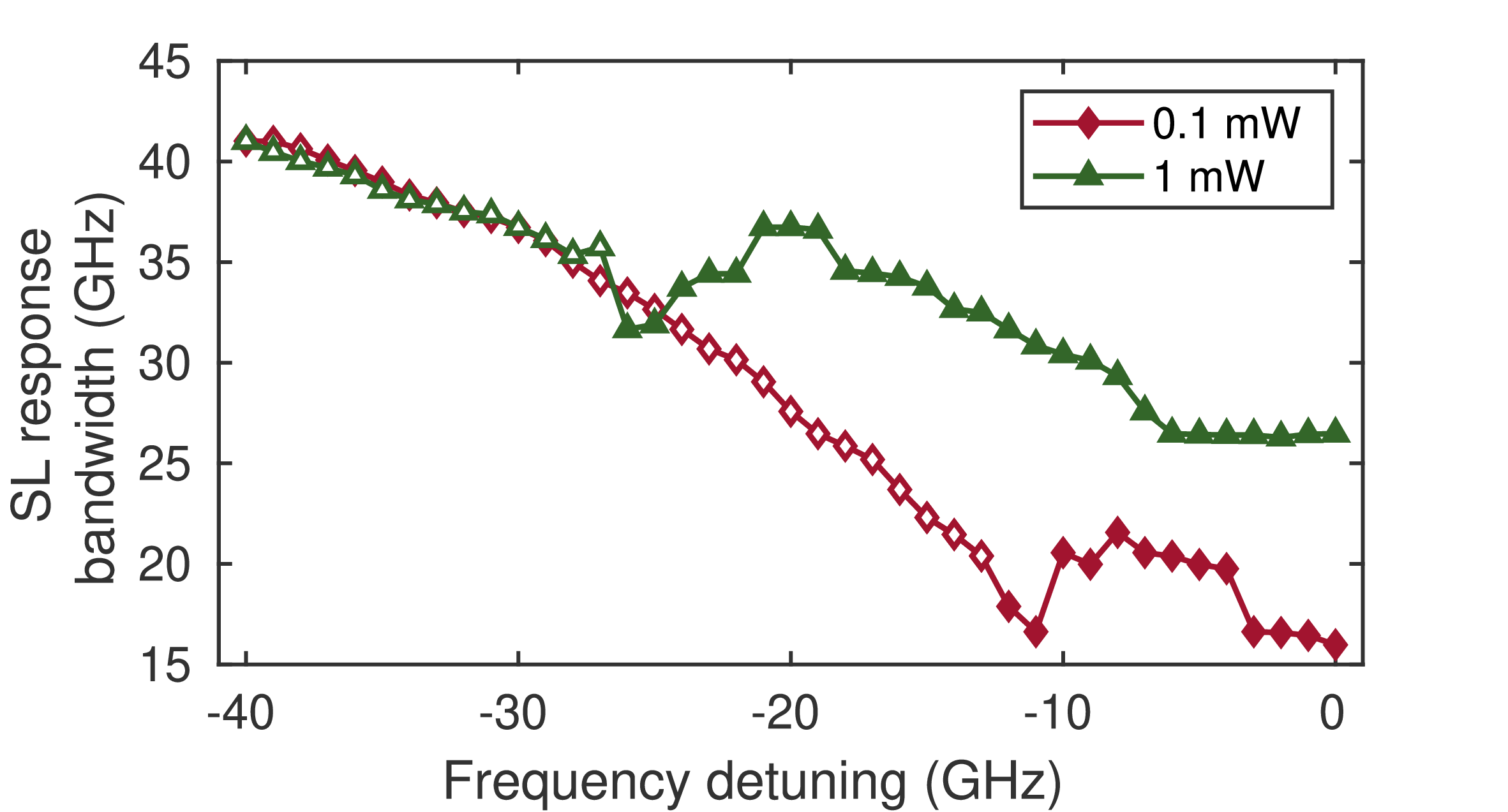}
\caption{Response bandwidth of the photonic TDRC. Empty symbols: partial locked or unlocked operation. Filled symbols: injection locked operation.}
\label{fig:BW}
\end{figure}

Fig. \ref{fig:BW} shows the response bandwidth of the photonic reservoir versus the frequency detuning between the drive and the reservoir laser ($\Delta f$), for two different levels of average optical injection power - 0.1 mW (red) and 1 mW (green) - and a frequency resolution of 1 GHz. 
If the lasers are injection locked (partially locked or unlocked), the points in this graph are color filled (empty). 
We consider that the lasers are partially locked or unlocked if we can identify amplified spectral components centered at an RF frequency equal to $\Delta f$.
When increasing the average optical injection power from 0.1 mW to 1 mW, we observe the following: first, there is an enhancement of the response bandwidth of the system of at least 5 GHz; second, the injection locking is observed for a wider region of the $\Delta f$ parameter space.
For both injection conditions, we observe a local dip in the response bandwidth, but for different $\Delta f$.
The dip emerges from a dynamical bistability region of the reservoir's response and is associated with the boundary between locking and unlocking \cite{herrera2021using}. 
The dip is located at $\Delta f = -12$ GHz for the case of 0.1 mW, and at $\Delta f = -25$ GHz for the case of 1 mW injection.
For lower $\Delta f$ than for the location of the dip, partial unlocking of the injection laser starts to appear, for both optical injection conditions. 
However, our calculation of the response bandwidth of the reservoir does not originate only from the bandwidth enhancement effect (empty symbols in Fig. \ref{fig:BW}).

\subsection{TDRC performance}
\begin{figure}[t]
\centering\includegraphics[width=0.85\columnwidth]{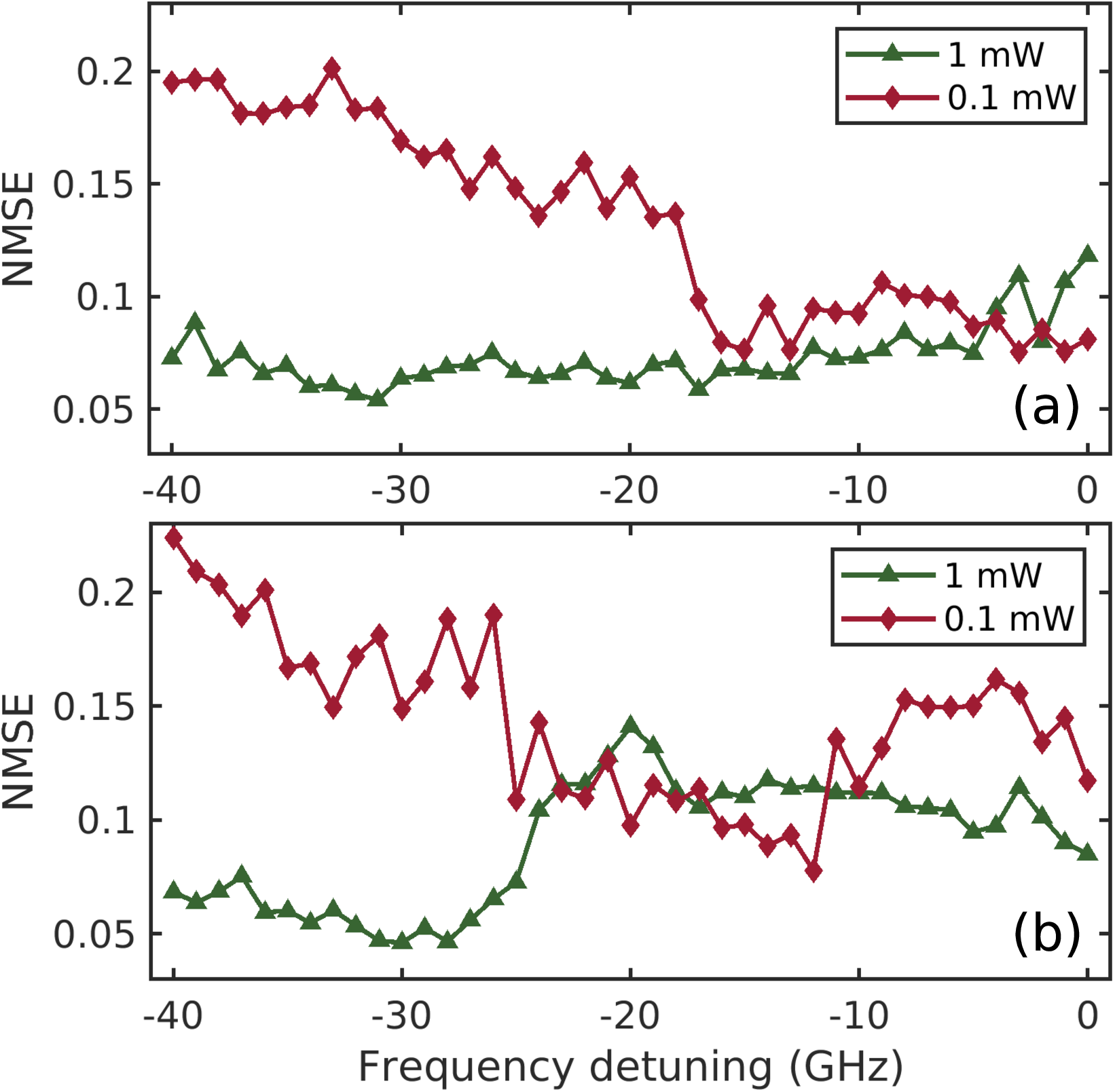}
\caption{NMSE in the Santa Fe one-step-ahead prediction task, for (a) $\theta_l =$ 93.75 ps, and (b) $\theta_s =$ 11.72 ps, and for two levels of average optical injection power: 0.1 mW (red) and 1 mW (green).} 
\label{fig:NMSE}
\end{figure}

Bandwidth enhancement is favorable for computing at faster rates. 
The final computing TDRC performance depends, however, also on other attributes that contribute to the nonlinear transformation between the input and output response.
Here, we explore the impact of injection strength and frequency detuning.
In Fig. \ref{fig:NMSE}, we show the computing performance for the Santa Fe one-step-ahead prediction task, for $\theta_l = 93.75$ ps and $\theta_s = 11.72$ ps, and for two levels of optical injection (0.1 mW and 1 mW), versus $\Delta f$. 
For $\theta_l = 93.75$ ps (Fig. \ref{fig:NMSE} a), when a moderate injection is considered (0.1 mW), we identify two regions ($\Delta f \sim 0$ GHz and $\Delta f \sim -12$ GHz) with the lowest values of the NMSE (0.076). 
When a strong injection is considered (1 mW), the lowest NMSE is 0.054 and achieved for $\Delta f = -31$ GHz. 
The smallest NMSE is found for a frequency detuning slightly lower than the one of the bistability region, independently of the injection level.
For these conditions, the lasers are partially locked, and the output response exhibits high consistency. 
The latter is calculated as the average consistency correlation ($C_{av}$) among 10 individual, non-averaged output responses of the reservoir.
For $\Delta f = -31$ GHz, $C_{av}$ is 0.85, while for injection locked operation ($\Delta f = 0$ GHz) is slightly lower ($C_{av}$= 0.82).
For $\theta_s = 11.72$ ps, we again observe a reduction of the NMSE for the strong injection case, as shown in Fig. \ref{fig:NMSE} b. 
Again here, the NMSE reaches its minimum value for $\Delta f$ slightly lower than the one of the bistability region.
The minima of the prediction error are obtained for $\Delta f = -12$ GHz, for the case of moderate injection, and $\Delta f = -31$ GHz, for the case of strong injection, with NMSE of 0.078 and 0.046, respectively. 
For more negative $\Delta f$, the effect of partial locking becomes progressively weaker until the lasers become completely unlocked. 
The SNR of the response signal decreases and the NMSE of the computing task increases significantly. 
Unlocked conditions are observed in Fig. \ref{fig:NMSE} for the case of moderate optical injection (0.1 mW), but not for the case of strong injection (1 mW) where the NMSE is low for values up to $\Delta$f = -40 GHz.

\begin{figure}[t]
\centering\includegraphics[width=0.85\columnwidth]{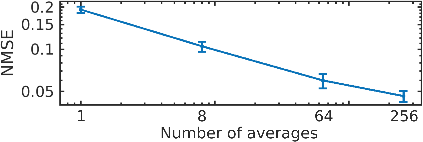}
\caption{NMSE performance of the Santa Fe computing task, for different levels of averaging of the output time series. Each point is obtained with 20 measurements.}
\label{fig:NMSE_av}
\end{figure}

Averaging the reservoir's output response has a significant impact on the NMSE performance (Fig. \ref{fig:NMSE_av}).
When considering $N$ = 256 averages, the SNR of the detected signal is improved by $N^{0.5}$.
Although some temporal dynamics of the reservoir's response are annulled by the averaging process, these seem to have a low contribution to the NMSE, as the variance of the error is very low when $N$ = 1 (Fig. \ref{fig:NMSE_av}).
It becomes clear that with averaging we do not address a real-time system.
However, we demonstrate that possible improvement of the SNR response of the conventional optical and optoelectronic components will allow an efficient real-time computing performance.

\subsection{TDRC dynamics}
\begin{figure*}[h]
  \centering\includegraphics[width=0.99\textwidth]{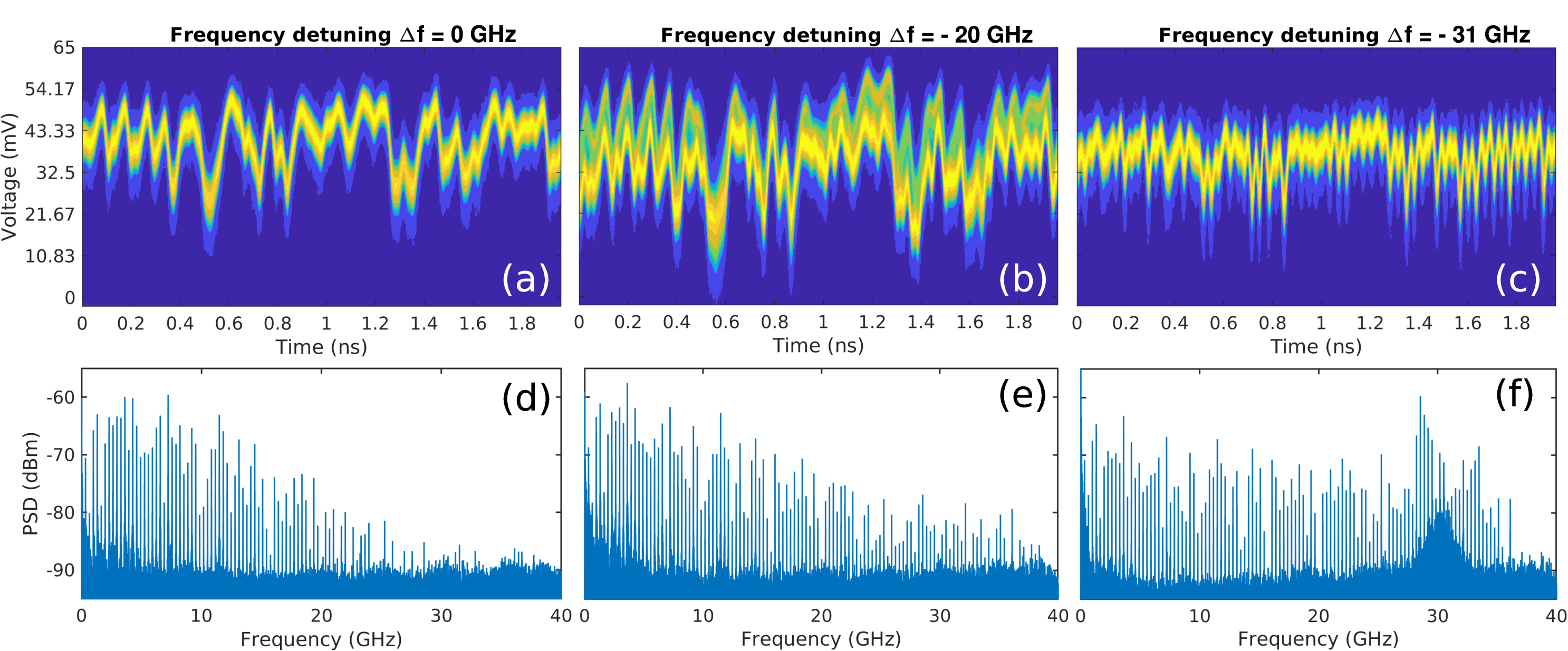}
  \caption{Dynamical response of the photonic reservoir after photodetection, for strong optical injection (1 mW) and $\theta_{s}$ = 12 ps, and for different $\Delta f$ conditions. a-c: Temporal persistence plots of the reservoir's response. d-f: The corresponding baseband power spectra. PSD: Power spectral density.}
  \label{fig:dynamics}
\end{figure*}

To gain a better understanding of the computing performance, we study the persistence plots of the reservoir's temporal response (Fig. \ref{fig:dynamics} a-c) and the corresponding spectra (Fig. \ref{fig:dynamics} d-f), for different $\Delta f$ conditions. 
The spectra are obtained with a 44 GHz bandwidth electrical spectrum analyzer (Keysight, EXA N9010B). 
For $\Delta f$ = 0 GHz, the response laser is injection-locked to the drive laser.
The obtained RF spectrum (Fig. \ref{fig:dynamics} d) consists of different frequency components that result from the characteristic time scales of the encoded masked information. The persistence plot (Fig. \ref{fig:dynamics} a) shows the system's response to a masked input of a 2 ns duration. 
When setting $\Delta f$ < 0 GHz, and in the presence of injection-locked operation, the system's response to the masked input also includes higher frequency components.
For $\Delta f = -20$ GHz, this is reflected in the RF spectrum (Fig. \ref{fig:dynamics} e), where frequencies above 30 GHz appear, due to the bandwidth enhancement of the response laser.
The persistence plot in this case (Fig. \ref{fig:dynamics} b) shows that the dynamical behavior of the response laser is affected by the underlying bistability that is associated with the locking and partial locking transition. 
The bistability results in time intervals with bimodal distributions of the intensity (e.g. time intervals of [0 ns, 0.5 ns] and [1.0 ns, 1.2 ns]).
For $\Delta f = -31$ GHz (Fig. \ref{fig:dynamics} c), the response laser is partially locked to the injection laser.
In the corresponding spectral distribution of Fig. \ref{fig:dynamics} f, we observe the appearance of high-power, spectral components centered at an RF frequency close to $\Delta f$. 
This is also reflected in the time series of Fig. \ref{fig:dynamics} c, with the existence of much faster oscillations.
This operating region, which combines high consistency and bandwidth-enhanced operation, exhibits the lowest prediction errors.

\newpage
\section{Conclusions}

We demonstrated experimentally the advantage of bandwidth-enhanced operation of a photonic TDRC, speeding up the computation without sacrificing performance. 
The induced fast nonlinear transient responses through bandwidth enhancement allowed our TDRC system to operate with a virtual node time separation of 11.72 ps - the smallest reported experimentally to our knowledge – while achieving a very low prediction error (NMSE = 0.046) for the nonlinear Santa Fe time series prediction task.
Therefore, even when being restricted to short delays, reasonable numbers of virtual nodes can still be implemented.
This opens further perspectives for the application of integrated semiconductor laser-based TDRC implementations.

\section*{Acknoledgements}
This work has been partially supported by the María de Maeztu project CEX2021-001164-M funded by the  MCIN/AEI/10.13039/501100011033. The work of I. Est\'ebanez has been supported by MICINN, AEI, FEDER and the University of the Balearic Islands through a predoctoral fellowship (MDM-2017-0711-18-2).
% or


%\section*{Acknowledgment}

\ifCLASSOPTIONcaptionsoff
  \newpage



%\bibliographystyle{IEEEtran}
\fi
\bibliographystyle{IEEEtran}
\bibliography{IEEEabrv,main}

\begin{thebibliography}{10}
\providecommand{\url}[1]{#1}
\csname url@samestyle\endcsname
\providecommand{\newblock}{\relax}
\providecommand{\bibinfo}[2]{#2}
\providecommand{\BIBentrySTDinterwordspacing}{\spaceskip=0pt\relax}
\providecommand{\BIBentryALTinterwordstretchfactor}{4}
\providecommand{\BIBentryALTinterwordspacing}{\spaceskip=\fontdimen2\font plus
\BIBentryALTinterwordstretchfactor\fontdimen3\font minus
  \fontdimen4\font\relax}
\providecommand{\BIBforeignlanguage}[2]{{%
\expandafter\ifx\csname l@#1\endcsname\relax
\typeout{** WARNING: IEEEtran.bst: No hyphenation pattern has been}%
\typeout{** loaded for the language `#1'. Using the pattern for}%
\typeout{** the default language instead.}%
\else
\language=\csname l@#1\endcsname
\fi
#2}}
\providecommand{\BIBdecl}{\relax}
\BIBdecl

\bibitem{shastri2021photonics}
B.~J. Shastri, A.~N. Tait, T.~Ferreira~de Lima, W.~H. Pernice, H.~Bhaskaran,
  C.~D. Wright, and P.~R. Prucnal, ``Photonics for artificial intelligence and
  neuromorphic computing,'' \emph{Nature Photonics}, vol.~15, no.~2, pp.
  102--114, 2021.

\bibitem{jaeger2004harnessing}
H.~Jaeger and H.~Haas, ``Harnessing nonlinearity: Predicting chaotic systems
  and saving energy in wireless communication,'' \emph{Science}, vol. 304, no.
  5667, pp. 78--80, 2004.

\bibitem{tanaka2019recent}
G.~Tanaka, T.~Yamane, J.~B. H{\'e}roux, R.~Nakane, N.~Kanazawa, S.~Takeda,
  H.~Numata, D.~Nakano, and A.~Hirose, ``Recent advances in physical reservoir
  computing: A review,'' \emph{Neural Networks}, vol. 115, pp. 100--123, 2019.

\bibitem{Nakajima2021}
\BIBentryALTinterwordspacing
K.~Nakajima and I.~Fischer, Eds., \emph{Reservoir Computing - Theory, Physical
  Implementations, and Applications}, ser. Natural Computing Series.\hskip 1em
  plus 0.5em minus 0.4em\relax Springer, 2021. [Online]. Available:
  \url{https://doi.org/10.1007/978-981-13-1687-6}
\BIBentrySTDinterwordspacing

\bibitem{appeltant2011information}
L.~Appeltant, M.~C. Soriano, G.~Van~der Sande, J.~Danckaert, S.~Massar,
  J.~Dambre, B.~Schrauwen, C.~R. Mirasso, and I.~Fischer, ``Information
  processing using a single dynamical node as complex system,'' \emph{Nature
  Communications}, vol.~2, no.~1, p. 468, 2011.

\bibitem{larger2012photonic}
L.~Larger, M.~C. Soriano, D.~Brunner, L.~Appeltant, J.~M. Guti{\'e}rrez,
  L.~Pesquera, C.~R. Mirasso, and I.~Fischer, ``Photonic information processing
  beyond turing: an optoelectronic implementation of reservoir computing,''
  \emph{Optics Express}, vol.~20, no.~3, pp. 3241--3249, 2012.

\bibitem{paquot2012optoelectronic}
Y.~Paquot, F.~Duport, A.~Smerieri, J.~Dambre, B.~Schrauwen, M.~Haelterman, and
  S.~Massar, ``Optoelectronic reservoir computing,'' \emph{Scientific Reports},
  vol.~2, no.~1, p. 287, 2012.

\bibitem{brunner2013parallel}
D.~Brunner, M.~C. Soriano, C.~R. Mirasso, and I.~Fischer, ``Parallel photonic
  information processing at gigabyte per second data rates using transient
  states,'' \emph{Nature Communications}, vol.~4, no.~1, p. 1364, 2013.

\bibitem{estebanez2020accelerating}
I.~Est{\'e}banez, J.~Schwind, I.~Fischer, and A.~Argyris, ``Accelerating
  photonic computing by bandwidth enhancement of a time-delay reservoir,''
  \emph{Nanophotonics}, vol.~9, no.~13, pp. 4163--4171, 2020.

\bibitem{stelzer2020performance}
F.~Stelzer, A.~R{\"o}hm, K.~L{\"u}dge, and S.~Yanchuk, ``Performance boost of
  time-delay reservoir computing by non-resonant clock cycle,'' \emph{Neural
  Networks}, vol. 124, pp. 158--169, 2020.

\bibitem{weigend1993results}
A.~Weigend and N.~Gershenfeld, ``Results of the time series prediction
  competition at the santa fe institute,'' in \emph{IEEE International
  Conference on Neural Networks}, 1993, pp. 1786--1793 vol.3.

\bibitem{ortin2020delay}
S.~Ort{\'\i}n and L.~Pesquera, ``Delay-based reservoir computing: tackling
  performance degradation due to system response time,'' \emph{Optics Letters},
  vol.~45, no.~4, pp. 905--908, 2020.

\bibitem{bueno2017conditions}
J.~Bueno, D.~Brunner, M.~C. Soriano, and I.~Fischer, ``Conditions for reservoir
  computing performance using semiconductor lasers with delayed optical
  feedback,'' \emph{Optics Express}, vol.~25, no.~3, pp. 2401--2412, 2017.

\bibitem{simpson1995bandwidth}
T.~Simpson, J.~Liu, and A.~Gavrielides, ``Bandwidth enhancement and broadband
  noise reduction in injection-locked semiconductor lasers,'' \emph{IEEE
  Photonics Technology Letters}, vol.~7, no.~7, pp. 709--711, 1995.

\bibitem{liu1997modulation}
J.~Liu, H.~Chen, X.~Meng, and T.~Simpson, ``Modulation bandwidth, noise, and
  stability of a semiconductor laser subject to strong injection locking,''
  \emph{IEEE Photonics Technology Letters}, vol.~9, no.~10, pp. 1325--1327,
  1997.

\bibitem{murakami2003cavity}
A.~Murakami, K.~Kawashima, and K.~Atsuki, ``Cavity resonance shift and
  bandwidth enhancement in semiconductor lasers with strong light injection,''
  \emph{IEEE Journal of Quantum Electronics}, vol.~39, no.~10, pp. 1196--1204,
  2003.

\bibitem{wang2008enhancing}
A.~Wang, Y.~Wang, and H.~He, ``Enhancing the bandwidth of the optical chaotic
  signal generated by a semiconductor laser with optical feedback,'' \emph{IEEE
  Photonics Technology Letters}, vol.~20, no.~19, pp. 1633--1635, 2008.

\bibitem{kanno2016complexity}
K.~Kanno, A.~Uchida, and M.~Bunsen, ``Complexity and bandwidth enhancement in
  unidirectionally coupled semiconductor lasers with time-delayed optical
  feedback,'' \emph{Physical Review E}, vol.~93, no.~3, p. 032206, 2016.

\bibitem{herrera2021using}
D.~J. Herrera, V.~Kovanis, and L.~F. Lester, ``Using transitional points in the
  optical injection locking behavior of a semiconductor laser to extract its
  dimensionless operating parameters,'' \emph{IEEE Journal of Selected Topics
  in Quantum Electronics}, vol.~28, no.~1, pp. 1--9, 2021.

\end{thebibliography}

\end{document}